\documentclass[11pt]{article}
\usepackage{jcappub,natbib,float,caption,comment,bm,amsmath,wasysym,subfig}
\usepackage{ulem}
\usepackage[usenames,dvipsnames]{xcolor}
\usepackage{hyperref}
\def\be{\begin{equation}}
\def\ee{\end{equation}}
\def\ba#1\ea{\begin{align}#1\end{align}}

\newcommand{\refeq}[1]{equation~(\ref{eq:#1})}
\newcommand{\refeqs}[2]{equations~(\ref{eq:#1})--(\ref{eq:#2})}
\newcommand{\refEq}[1]{Equation~(\ref{eq:#1})}

\newcommand{\reffig}[1]{figure~\ref{fig:#1}}

\newcommand{\refFig}[1]{Figure~\ref{fig:#1}}
\newcommand{\refsec}[1]{section~\ref{sec:#1}}

\newcommand{\refapp}[1]{appendix~\ref{app:#1}}

\newcommand{\pd}{\partial}

\title{Non-Gaussianity of Inflationary Gravitational Waves from the Field Equation}
\author[a]{Aniket Agrawal}
\affiliation[a]{Max-Planck-Institut f\"ur Astrophysik, Karl-Schwarzschild-Str. 1, 85748 Garching, Germany}
\emailAdd{aniket@mpa-garching.mpg.de}
\abstract{
We demonstrate equivalence of the in-in formalism and Green's function method for calculating the bispectrum of primordial gravitational waves generated by vacuum fluctuations of the metric. The tree-level bispectrum from the field equation, $B_h$, agrees with the results obtained previously using the in-in formalism exactly. Characterising non-Gaussianity of the fluctuations using the ratio $B_h/P^2_h$ in the equilateral configuration, where $P_h$ is the power spectrum of scale-invariant gravitational waves,  we show that it is much weaker than in models with spectator gauge fields. We also calculate the tree-level bispectrum of two right-handed and one left-handed gravitational wave using Green's function, reproducing the results from in-in formalism, and show that it can be as large as the bispectrum of three right-handed gravitational waves.   
}

\begin{document}

\maketitle
\flushbottom

\section{Introduction}
\label{sec:intro}

Vacuum fluctuations of the metric tensor during inflation are weakly non-Gaussian~\cite{Salopek:1990,Falk:1992sf,Gangui:1993tt}. This non-Gaussianity arises from self-interactions in (non-linear) general relativity, which make Gaussian vacuum fluctuations slightly non-Gaussian. Self-interactions among scalar fluctuations of the metric tensor leave their imprints on non-Gaussianity of temperature anisotropies of the cosmic microwave background (CMB). WMAP~\cite{hinshaw/etal:2013} and Planck~\cite{planckNG:2015} have used non-Gaussianity of temperature fluctuations extensively to put constraints on different models of inflation, allowing us to probe non-linearities and interactions of fields in the primordial Universe. Tensor fluctuations, on the other hand, leave their imprint on B-mode polarisation of the CMB~\cite{seljak/zaldarriaga:1996,kamionkowski/kosowsky/stebbins:1996}. If these originate from vacuum fluctuations of the metric, they are also expected to be weakly non-Gaussian~\cite{maldacena:2002,maldacena/pimentel:2011,Gao:2011vs}. However, they can also originate from other sources, such as scalar fields~\cite{cook/sorbo:2012,carney/etal:2012,biagetti/fasiello/riotto:2013,senatore/silverstein/zaldarriaga:2014}, U(1) gauge
fields~\cite{sorbo:2011,anber/sorbo:2012,barnaby/peloso:2011,barnaby/etal:2012,peloso/sorbo/unal:2016}, and SU(2) gauge
fields~\cite{maleknejad/sheikh-jabbari:2013,dimastrogiovanni/peloso:2012,adshead/etal:2013,adshead/martinec/wyman:2013,maleknejad:2016,dimastrogiovanni/fasiello/fujita:2016}. In these cases, they can be  highly non-Gaussian, producing a large bispectrum~\cite{cook/sorbo:2013, namba/etal:2015,Agrawal:2017awz,Agrawal:2018mrg}. Consequently, non-Gaussianity of tensor fluctuations is a definitive test of their vacuum or sourced origin. It is therefore desirable to have a method to compute non-Gaussianity easily for any model of our choice.

Most computations of non-Gaussianity are carried out using the in-in formalism~\cite{Schwinger:1960qe,Bakshi:1962dv,Keldysh:1964ud,Chou:1984es,Jordan:1986ug,Calzetta:1986ey,Weinberg:2005vy}, which has been inspired by methods of high-energy physics and quantum field theory. There, one starts with the interaction Hamiltonian, and uses the interaction picture, where operators are evolved freely while states are evolved using the interaction Hamiltonian. Non-Gaussian correlators of the field are then evaluated as expectation values of commutators of the correlators with the interaction Hamiltonian, in the ``in" state corresponding to the vacuum of the interacting theory. The tree-level tensor bispectrum for vacuum fluctuations has also been calculated previously using this method~\cite{maldacena:2002,maldacena/pimentel:2011,Gao:2011vs}. While the in-in formalism is quite useful for calculation of correlation functions, allowing one to carry over methods and intuition from quantum field theory into cosmology, there is still considerable merit in exploring alternative methods. Other techniques can allow calculation of non-Gaussianity for a wider range of models or provide technical simplifications in the computation~\cite{seery:2008}. Green's function method provides one such alternative.

Traditionally, Green's function method was used to calculate the tree-level spectrum of a scalar field in de Sitter spacetime for the first time~\cite{Birrell:1982ix}, as well as to calculate the tree-level spectrum of the curvature perturbation using the Mukhanov-Sasaki equation~\cite{Mukhanov:1985rz,Sasaki:1986hm}. More recently, it was applied to calculations of non-Gaussianity of scalar field fluctuations~\cite{seery:2008}. It has also been used previously to calculate the tensor bispectrum in models with sources~\cite{barnaby/peloso:2011,barnaby/etal:2012,cook/sorbo:2013,namba/etal:2015,Agrawal:2017awz}. The key advantage of Green's functions shows up in calculation of 1-loop (or higher) order diagrams. While in the in-in formalism one needs to calculate second-order commutators to obtain 1-loop spectra, in the Green's function approach one can use the second-order operator directly, once it has been obtained to calculate tree-level spectra. Moreover, as pointed out in~\cite{seery:2008}, the method of field equations adopted in this paper enables one to calculate non-Gaussianity in a model even if it is not descended from an effective action principle. Furthermore, Green's function methods are common in classical mechanics, and extending them to calculate non-Gaussianity of quantum fields merely involves promoting fields to operators and invoking commutation relations. Thus, it provides a simple and easy recipe to calculate the bispectrum. 

The rest of the paper is organised as follows : in \refsec{powerspectrum} we derive the equation of motion of tensor fluctuations up to second-order and review the calculation of the power spectrum in de Sitter space to set up notation. Then in \refsec{bispectrum} we use the Green's function method to calculate the tree-level bispectrum of three right-handed tensor fluctuations, and show that we recover the result derived in~\cite{maldacena:2002,maldacena/pimentel:2011,Gao:2011vs} using the in-in formalism if the same initial vacuum is chosen in the Green's function approach. In \refsec{bispectrumRRL} we derive the tree-level bispectrum of two right-handed and one left-handed tensor fluctuation using Green's functions (which we refer to as the ``mixed" bispectrum) and show that it can be as large as that of three right-handed fluctuations. A brief conclusion is provided in  \refsec{conclusion}. The derivation of the source function at second-order is presented in \refapp{source}.

\section{Vacuum Fluctuations of the Tensor Metric}
\label{sec:powerspectrum}

We start with a flat Friedmann-Lema\^itre-Robertson-Walker (FLRW) metric as the background, written in conformal time, $x^0 \equiv \tau$, and comoving space, $x^i$, co-ordinates : $ds^2 = g_{\mu\nu}dx^{\mu}dx^{\nu}$ with $g_{00}=a^2$, $g_{0i}=g_{i0}=0$, $g_{ij}=-a^2\delta_{ij}$, where $a\equiv a(\tau)$ is the scale factor of the Universe. We also assume de Sitter expansion throughout so that $a(\tau) = -1/H\tau$, where $H$ is the Hubble parameter during inflation. Tensor perturbations of the metric are then written as
\begin{equation}
g_{ij} = -a^2(\delta_{ij}+h_{ij})\,,
\end{equation}
where we impose transverse and traceless conditions on $h_{ij}$, $\delta^{ij}h_{ij}=0=\partial_i h_{ij} = \partial_j h_{ij}$. The inverse metric is given by $g^{ij} = -a^{-2}(\delta^{ij}-h^{ij}+h^{ik}h^{kj}+\mathcal{O}(h^3))$. We also define a canonically normalised tensor perturbation, $\psi_{ij}$, as
\begin{equation}\label{eq:psi_defn}
\psi_{ij} = \frac{a M_P}{2} h_{ij}=-\frac{1}{2\tau}\frac{M_P}{H}h_{ij}\,,
\end{equation}
where $M_P \equiv 1/\sqrt{8\pi G}$ is the reduced Planck mass. We assume Einstein gravity, so that the field equations can be derived by varying the Einstein-Hilbert action,
\begin{equation}\label{eq:gr_action}
S = \int d\tau d^3x \sqrt{-g}\Bigg[\frac{M^2_P}{2}R+\mathcal{L}_m\Bigg]\,,
\end{equation}
where $g$ is the determinant of the metric tensor, $R$ is the Ricci scalar, and $\mathcal{L}_m$ is the Lagrangian density of energy-momentum sources in the Universe. Since we are only interested in vacuum fluctuations of the metric we set $\mathcal{L}_m = 0$. 

To obtain the power spectrum and bispectrum of GWs we need to expand the action, \refeq{gr_action}, up to third order in perturbations, $S = S^{(2)}+S^{(3)}$, where the second-order action is given as~\cite{maldacena:2002}
\begin{equation}\label{eq:s2}
S^{(2)} = \int d\tau d^3x \Bigg[\frac{1}{2}\psi'_{ij}\psi'_{ij}-\frac{1}{2}\pd_k \psi_{ij}\pd_k \psi_{ij}+\frac{1}{\tau^2}\psi_{ij}\psi_{ij}\Bigg]\,,
\end{equation}
where prime denotes derivative w.r.t. the conformal time $\tau $, and $\pd_k$ denotes $\pd/\pd x^k$. The third-order action is given as~\cite{maldacena:2002}
\begin{equation}\label{eq:s3}
S^{(3)} = -\int d\tau d^3x \, \frac{2H}{M_P}\tau\Big[\pd_l\psi_{ik}\pd_k\psi_{ij}\psi_{jl}-\frac{1}{2}\pd_l\psi_{ij}\pd_k\psi_{ij}\psi_{kl}\Big]\,.
\end{equation}
The quadratic and cubic actions, \refeqs{s2}{s3}, then give the equations of motion for GWs as (note that it is possible to obtain this equation directly by perturbing the Einstein equation without invoking the Einstein-Hilbert action at all)
\begin{multline}\label{eq:h_eom3}
\psi_{pq}''(\tau, \bm x)-\Bigg(\pd_l^2+\frac{2}{\tau^2}\Bigg)\psi_{pq}(\tau, \bm x)=\frac{H}{M_P}\,\tau\,\Big[2\psi_{pl,qr}(\tau, \bm x)\psi_{lr}(\tau, \bm x)+2\psi_{pr,l}(\tau, \bm x)\psi_{ql,r}(\tau, \bm x)
\\
+\psi_{lr,q}(\tau, \bm x)\psi_{lr,p}(\tau, \bm x)-2\psi_{pq,lr}(\tau, \bm x)\psi_{lr}(\tau, \bm x)-2\psi_{lr,q}(\tau, \bm x)\psi_{lp,r}(\tau, \bm x)\Big]\,.
\end{multline}
To proceed further, we decompose $\psi_{ij}$ into a left- and right-handed component, using the circular polarisation tensors,
\begin{equation}\label{eq:psi_decomp}
\psi_{ij}(\tau, \bm{x}) = \int \frac{d^3k}{(2\pi)^3} e^{i\bm{k}\cdot \bm{x}} \Big[e^R_{ij}(\bm{k})\psi_{\bm{k}}^R(\tau)+e^L_{ij}(\bm{k})\psi_{\bm{k}}^L(\tau)\Big]\,,
\end{equation}
where $e^p_{ij}\,(p=L,R)$ are polarisation tensors\footnote{For a summary of their properties, please see Appendix A of~\cite{Agrawal:2018mrg}.}. Note that we normalise $e^p_{ij}$ such that $e^R_{ij}(\bm{k})e^R_{ij}(-\bm{k})=e^L_{ij}(\bm{k})e^L_{ij}(-\bm{k})=1$. Next we quantize the field $\psi^p_{\bm{k}}$ and expand it in a perturbative series as~\cite{seery:2008}
\begin{equation}
\hat{\psi}_{\bm{k}}^p(\tau) = \hat{\psi}^p_1(\tau, \bm{k})+\hat{\psi}^p_2(\tau, \bm{k})+\ldots\,.
\end{equation}
This expansion is useful when we identify $\hat{\psi}^p_1(\tau, \bm{k})$ with the part of the field that obeys a linear equation of motion, $\hat{\psi}^p_2(\tau, \bm{k})$ obeys a second-order equation, and so on. It then follows that $\hat{\psi}^p_1(\tau, \bm{k})$ obeys Gaussian statistics exactly~\cite{grishchuk:1974,starobinsky:1979,starobinsky:1985}, and all higher order fields vanish at past infinity~\cite{seery:2008}. The first-order component is written as
\begin{equation}\label{eq:ann_cre}
\hat{\psi}^p_1(\tau, \bm{k}) = \Psi^p_1(\tau, k)\hat{a}^p_{\bm{k}}+\Psi^{*p}_1(\tau, k)\hat{a}^{\dagger p}_{-\bm{k}}\,,
\end{equation}
where $\Psi^p_1(\tau, k)$ represents the linear mode function, and $\hat{a}^{\dagger p}_{\bm{k}}$ and $\hat{a}^p_{\bm{k}}$ represent the creation and annihilation operators for modes with momentum $\bm{k}$. They satisfy the canonical commutation relations $[\hat{a}^{p}_{\bm{k}}, \hat{a}^{\dagger q}_{-\bm{k}'}] = (2\pi)^3 \delta_D(\bm{k}+\bm{k}')\delta^{pq}$. The mode function $\Psi^p_1(\tau, k)$, that exhibits Gaussian statistics, quadratically sources $\hat{\psi}^p_2(\tau, \bm{k})$ through the second-order terms in \refeq{h_eom3}, making $\hat{\psi}^p_2(\tau, \bm{k})$ non-Gaussian. 

For the linear mode functions the equation of motion in Fourier space yields~\cite{grishchuk:1974,starobinsky:1979}
\begin{equation}\label{eq:eom_lin}
\pd_y^2 \Psi^p_1 + \Bigg[1-\frac{2}{y^2}\Bigg]\Psi^p_1 = 0\,,
\end{equation}
where $y\equiv -k\tau$. Note that the two polarisations $p$, independently satisfy \refeq{eom_lin}. Thus vacuum fluctuations produce parity invariant GWs at first-order. The solution for this equation can be analytically calculated and is given by
\begin{equation}\label{eq:psi_1}
\Psi^p_1(\tau, k) = \frac{e^{-ik\tau}}{\sqrt{2k}}\Bigg[1-\frac{i}{k\tau}\Bigg]
\end{equation}
for each value of $p$, where initial conditions are set by requiring that as $-k\tau \rightarrow \infty$ the mode function should reduce to the Minkowsi space mode function, $e^{-ik\tau}/\sqrt{2k}$. Using \refeq{psi_1} and \refeq{psi_defn} we find
\begin{equation}\label{eq:h_1}
h^p_1(\tau, k) = i\frac{H}{M_P}\sqrt{\frac{2}{k^3}}e^{-ik\tau}[1+{i}{k\tau}]\,,
\end{equation}
from which we get the power spectrum for each polarisation of GWs in the super-horizon limit $(k\tau \rightarrow 0)$ as
\begin{equation}\label{eq:ph}
P^{L/R}_h(k) = \Bigg(\frac{H}{M_P}\Bigg)^2\frac{2}{k^3} \quad \Rightarrow\, \mathcal{P}^{L/R}_h(k) \equiv \frac{k^3 P^{L/R}_h}{2\pi^2} = \Bigg(\frac{H}{M_P}\Bigg)^2\frac{1}{\pi^2}\,,
\end{equation}
where $\mathcal{P}^{L/R}_h$ is the dimensionless power spectrum of $h^{L/R}_1$, which is found to be scale-independent. At first-order, the two polarisations are uncorrelated with each other so that
\begin{equation}
\mathcal{P}^T_h = \mathcal{P}^L_h+\mathcal{P}^R_h = \Bigg(\frac{H}{M_P}\Bigg)^2\frac{2}{\pi^2}\,.
\end{equation}

\section{Bispectrum of Vacuum Tensor Fluctuations}
\label{sec:bispectrum_master}

Let us now consider the second-order terms in \refeq{h_eom3}. Although in first-order, the two polarisations obey independent equations of motion, each of which depends only on their respective polarisations, the same is not true in second-order. To see this more explicitly, we substitute \refeq{psi_decomp} into the left hand side of \refeq{h_eom3}, and write it in Fourier space,
\begin{multline}\label{eq:h_eom4}
e^R_{pq}(\bm{k})\Bigg[\psi^{''R}_{\bm{k}}+\Bigg(k^2-\frac{2}{\tau^2}\Bigg)\psi^R_{\bm{k}}\Bigg]+e^L_{pq}(\bm{k})\Bigg[\psi^{''L}_{\bm{k}}+\Bigg(k^2-\frac{2}{\tau^2}\Bigg)\psi^L_{\bm{k}}\Bigg] =S_{pq}(\bm{k})\,,
\end{multline}
where $S_{pq}(\bm{k})$ is the Fourier transform of the source function on the right hand side of \refeq{h_eom3}. The full expression for $S_{pq}(\bm{k})$ is given in \refapp{source}, which is indeed seen to depend symmetrically on first-order mode functions of both polarisations. 

The equation of motion for either polarisation is found by multiplying both sides of \refeq{h_eom4} by the polarisation tensor of the opposite helicity and summing over the indices $p$ and $q$. So, the right-handed polarisation obeys 
\begin{equation}\label{eq:hr_eom}
\psi^{''R}_{\bm{k}}+\Bigg(k^2-\frac{2}{\tau^2}\Bigg)\psi^R_{\bm{k}}=e^L_{pq}(\bm{k})S_{pq}(\bm{k})\,.
\end{equation}
Note that, under a parity transformation, the left-handed polarisation obeys the same equation with the same source function, so that parity is preserved even at second-order. Promoting fields to operators and using Green's function, the solution to the above equation is written as
\begin{equation}\label{eq:psi_2nd_gr}
\hat{\psi}^R_{2}(\tau,\bm k) = \int_{-\infty}^{0}\, d\eta\, G_{\psi}^{(2)}(\tau, \eta, k) \hat{S}_{pq}(\eta, \bm k) e^L_{pq}(\bm k)\,.
\end{equation}
where $G_{\psi}^{(2)}(\tau, \eta, k)$ is Green's function for the second-order perturbation and is given as
\begin{equation}\label{eq:gpsi}
G_{\psi}^{(2)}(\tau,\eta,k) = \frac{\Theta(\tau-\eta)}{k^3\tau\eta}\Big[k(\eta-\tau)\cos(k(\tau-\eta))+(1+k^2\tau\eta)\sin(k(\tau-\eta))\Big]
\equiv \frac{\tilde{G}^{(2)}_{\psi}(\tau,\eta,k)}{k^3\tau\eta}\,,
\end{equation}
where $\Theta(x)$ is the Heaviside step function, and we define the numerator as another function $\tilde{G}^{(2)}_{\psi}$. $\hat{S}_{pq}$ is a function of two first-order tensor perturbations, $\hat{\psi}_{1}$, (c.f. \refapp{source})
\begin{equation}\label{eq:source}
\hat{S}_{pq}(\eta, \bm k) = \frac{H}{M_P}\int \frac{d^3p_1 d^3p_2}{(2\pi)^6}  \delta_D(\bm p_1 +\bm p_2-\bm k) \,\eta\, \sum_{c_1, c_2 = L,R}[\hat{\psi}^{c_1}_1(\eta,{\bm{p}_1})\hat{\psi}^{c_2}_1(\eta,\bm{p}_2) Q^{c_1c_2}_{pq}(\bm p_1, \bm p_2)]\,,
\end{equation}
where $c_i$ denote polarisation states of the tensor perturbations. 

\subsection{Bispectrum of Three Right-Handed Tensor Fluctuations}
\label{sec:bispectrum}

The tree-level bispectrum of right-handed tensor modes is then given by
\begin{multline}\label{eq:bk_parts}
\left\langle \hat{\psi}^R(\tau, \bm k_1)\hat{\psi}^R(\tau, \bm k_2)\hat{\psi}^R(\tau, \bm k_3)\right\rangle = \left\langle \hat{\psi}_1^R(\tau, \bm k_1)\hat{\psi}_1^R(\tau, \bm k_2)\hat{\psi}_2^R(\tau, \bm k_3)\right\rangle
\\
+\left\langle \hat{\psi}_1^R(\tau, \bm k_1)\hat{\psi}_2^R(\tau, \bm k_2)\hat{\psi}_1^R(\tau, \bm k_3)\right\rangle+\left\langle \hat{\psi}_2^R(\tau, \bm k_1)\hat{\psi}_1^R(\tau, \bm k_2)\hat{\psi}_1^R(\tau, \bm k_3)\right\rangle\,.
\end{multline}
Let us work out one of these terms in detail. Using \refeq{psi_2nd_gr}, we find
\begin{multline}\label{eq:bk_raw}
\left\langle \hat{\psi}_1^R(\tau, \bm k_1)\hat{\psi}_1^R(\tau, \bm k_2)\hat{\psi}_2^R(\tau, \bm k_3)\right\rangle  = \frac{H}{M_P}\int_{-\infty}^{0}\, {d\eta}\,{\eta}\, G_{\psi}^{(2)}(\tau, \eta, k_3)\int\, \frac{d^3p_1 d^3p_2}{(2\pi)^6} \, 
\\
\delta_D(\bm p_1 +\bm p_2-\bm k_3) \,\, Q^{RR}_{pq}(\bm p_1, \bm p_2) e^L_{pq}(\bm k_3) \left\langle \hat{\psi}^R_1(\tau,{\bm{k}_1})\hat{\psi}^R_1(\tau,{\bm{k}_2})\hat{\psi}^R_1(\eta,{\bm{p}_1})\hat{\psi}^R_1(\eta,\bm{p}_2)\right\rangle\,,
\end{multline}
where we only use $Q^{RR}_{pq}$ because there are only right-handed first-order operators in the expectation value (see \refapp{source} for more details). The expectation value is calculated as
\begin{align}\label{eq:exp_val}
\nonumber\left\langle \hat{\psi}^R_1(\tau,{\bm{k}_1})\hat{\psi}^R_1(\tau,{\bm{k}_2})\hat{\psi}^R_1(\eta,{\bm{p}_1})\hat{\psi}^R_1(\eta,\bm{p}_2)\right\rangle &= \left\langle \hat{a}^R_{\bm{k}_1}\hat{a}^R_{\bm{k}_2}\hat{a}^{R\dagger}_{-\bm{p}_1}\hat{a}^{R\dagger}_{-\bm{p}_2}\right\rangle 
\\
&\Psi^R_1(\tau,\bm{k}_1)\Psi^R_1(\tau,\bm{k}_2)\Psi^{*R}_1(\eta,\bm{p}_1)\Psi^{*R}_1(\eta,\bm{p}_2)
\\
\nonumber&=(2\pi)^6 \Big[\delta_{\bm{k}_1\bm{p}_1}\delta_{\bm{k}_2\bm{p}_2}+\delta_{\bm{k}_2\bm{p}_1}\delta_{\bm{k}_2\bm{p}_1}\Big]
\\
&\Psi^R_1(\tau,\bm{k}_1)\Psi^R_1(\tau,\bm{k}_2)\Psi^{*R}_1(\eta,\bm{p}_1)\Psi^{*R}_1(\eta,\bm{p}_2)\,,
\end{align} 
where we have defined $\delta_{\bm{k}_1\bm{p}_1}\equiv \delta_D(\bm{k}_1+\bm{p}_1)$. Substituting this expectation value in \refeq{bk_raw} and integrating over the delta functions, we get
\begin{multline}\label{eq:bk_konly}
\left\langle \hat{\psi}_1^R(\tau, \bm k_1)\hat{\psi}_1^R(\tau, \bm k_2)\hat{\psi}_2^R(\tau, \bm k_3)\right\rangle  = (2\pi)^3 \delta_D(\bm k_1+\bm k_2+\bm k_3) \frac{H}{M_P}
\\
\int_{-\infty}^{0}\, {d\eta}\,{\eta}\, G_{\psi}^{(2)}(\tau, \eta, k_3) \Big[e^L_{pq}(\bm{k}_3)Q^{RR}_{pq}(-\bm{k}_1,-\bm{k}_2)+e^L_{pq}(\bm{k}_3)Q^{RR}_{pq}(-\bm{k}_2,-\bm{k}_1)\Big]
\\
\Psi^R_1(\tau,\bm{k}_1)\Psi^R_1(\tau,\bm{k}_2)\Psi^{*R}_1(\eta,-\bm{k}_1)\Psi^{*R}_1(\eta,-\bm{k}_2)\,.
\end{multline}
The contraction of the polarisation tensors is given by (\refapp{source})
\begin{equation}
\Big[e^L_{pq}(\bm{k}_3)Q^{RR}_{pq}(-\bm{k}_1,-\bm{k}_2)+e^L_{pq}(\bm{k}_3)Q^{RR}_{pq}(-\bm{k}_2,-\bm{k}_1)\Big] = k^2_1\, \tilde{\Xi}^2\, \Xi\,,
\end{equation}
where 
\begin{equation}\label{eq:xi_box}
\tilde{\Xi} = 1+r_2+r_3\,,\quad \Xi = \frac{(1+r_2+r_3)^3}{64r^2_2r^2_3}(r_2+r_3-1)(r_2-r_3+1)(-r_2+r_3+1)\,,
\end{equation}
with $r_2 \equiv k_2/k_1$ and $r_3 = k_3/k_1$~\cite{Agrawal:2018mrg}. We find that this contraction remains the same regardless of the order in which the different $\bm{k}$'s appear, and so is common to all three contributions to the bispectrum in \refeq{bk_parts}. Note that it is also independent of the conformal time, so we can take it out of the $\eta$ integral. This gives us
\begin{multline}\label{eq:bk_contracted}
\left\langle \hat{\psi}_1^R(\tau, \bm k_1)\hat{\psi}_1^R(\tau, \bm k_2)\hat{\psi}_2^R(\tau, \bm k_3)\right\rangle  = (2\pi)^3 \delta_D(\bm k_1+\bm k_2+\bm k_3) \frac{H}{M_P} k^2_1\,\tilde{\Xi}^2\,\Xi
\\
\Psi^R_1(\tau,\bm{k}_1)\Psi^R_1(\tau,\bm{k}_2)\int_{-\infty}^{0}\, {d\eta}\,{\eta}\, G_{\psi}^{(2)}(\tau, \eta, k_3) \Psi^{*R}_1(\eta,-\bm{k}_1)\Psi^{*R}_1(\eta,-\bm{k}_2)\,.
\end{multline}
We now substitute the first-order mode functions, \refeq{psi_1}, into the above equation to get
\begin{align}\label{eq:int1}
\nonumber&\Psi^R_1(\tau,\bm{k}_1)\Psi^R_1(\tau,\bm{k}_2)\int_{-\infty}^{0}\, {d\eta}\,{\eta}\, G_{\psi}^{(2)}(\tau, \eta, k_3) \Psi^{*R}_1(\eta,-\bm{k}_1)\Psi^{*R}_1(\eta,-\bm{k}_2) = 
\\
&\frac{e^{-ik_1\tau}}{\sqrt{2k_1}}\Big[1-\frac{i}{k_1\tau}\Big]\frac{e^{-ik_2\tau}}{\sqrt{2k_2}}\Big[1-\frac{i}{k_2\tau}\Big] \int_{-\infty}^{0}\, {d\eta}\,{\eta}\, G_{\psi}^{(2)}(\tau, \eta, k_3) \frac{e^{ik_1\eta}}{\sqrt{2k_1}}\Big[1+\frac{i}{k_1\eta}\Big]\frac{e^{ik_2\eta}}{\sqrt{2k_2}}\Big[1+\frac{i}{k_2\eta}\Big]\,.
\end{align}
Taking $-{i}/{k_j\tau}$ and $-{i}/{k_j\eta}$ out from the mode functions we get
\begin{align}
\nonumber&\Psi^R_1(\tau,\bm{k}_1)\Psi^R_1(\tau,\bm{k}_2)\int_{-\infty}^{0}\, {d\eta}\,{\eta}\, G_{\psi}^{(2)}(\tau, \eta, k_3) \Psi^{*R}_1(\eta,-\bm{k}_1)\Psi^{*R}_1(\eta,-\bm{k}_2) = 
\\
&\frac{e^{-i{(k_1+k_2)}\tau}}{4(k_1 k_2)^3\tau^2}[1+{i}{k_1\tau}][1+{i}{k_2\tau}] \int_{-\infty}^{0}\, {d\eta}\,{\eta}\, G_{\psi}^{(2)}(\tau, \eta, k_3) 
\frac{e^{i{(k_1+k_2)}\eta}}{\eta^2}[1-{i}{k_1\eta}][1-{i}{k_2\eta}]\,.
\end{align} 
Using \refeq{gpsi}, we can substitute Green's function,
\begin{align}
\nonumber&\Psi^R_1(\tau,\bm{k}_1)\Psi^R_1(\tau,\bm{k}_2)\int_{-\infty}^{0}\, {d\eta}\,{\eta}\, G_{\psi}^{(2)}(\tau, \eta, k_3) \Psi^{*R}_1(\eta,-\bm{k}_1)\Psi^{*R}_1(\eta,-\bm{k}_2) = 
\\
&\frac{e^{-i{(k_1+k_2)}\tau}}{4(k_1 k_2k_3)^3\tau^3}[1+{i}{k_1\tau}][1+{i}{k_2\tau}] \int_{-\infty}^{0}\, \frac{d\eta}{\eta^2}\, \tilde{G}_{\psi}^{(2)}(\tau, \eta, k_3) 
\frac{e^{i{(k_1+k_2)}\eta}}{\eta^2}[1-{i}{k_1\eta}][1-{i}{k_2\eta}]\,.
\end{align} 
Then the bispectrum of GWs is given as
\begin{align}\label{eq:bk_tau}
\nonumber&\left\langle \hat{h}_1^R(\tau, \bm k_1)\hat{h}_1^R(\tau, \bm k_2)\hat{h}_2^R(\tau, \bm k_3)\right\rangle  = (2\pi)^3 \delta_D(\bm k_1+\bm k_2+\bm k_3) \Bigg(\frac{H}{M_P}\Bigg)^4 k^2_1\,\tilde{\Xi}^2\,\Xi
\\
&\frac{2\,e^{-i{(k_1+k_2)}\tau}}{(k_1 k_2k_3)^3}[1+{i}{k_1\tau}][1+{i}{k_2\tau}] \int_{-\infty}^{0}\, \frac{d\eta}{\eta^2}\, \tilde{G}_{\psi}^{(2)}(\tau, \eta, k_3) 
{e^{i{(k_1+k_2)}\eta}}[1-{i}{k_1\eta}][1-{i}{k_2\eta}]\,,
\end{align} 
where we have used $\psi_{ij}(\tau,\bm{k}) \equiv aM_Ph_{ij}(\tau,\bm{k})/2$ (\refeq{psi_defn}). \refEq{bk_tau} is valid for any time $\tau$. However, to make contact with observations we need to take the super-horizon limit, $k_i \tau \rightarrow 0$, for all three modes. The function $\tilde{G}_{\psi}$ in this limit becomes (\refeq{gpsi})
\begin{equation}
\tilde{G}^{(2)}_{\psi}(0,\eta,k) = {\Theta(-\eta)}[k\eta\cos(k\eta)-\sin(k\eta)]\,,
\end{equation}
which we can re-write as
\begin{equation}\label{eq:gpsi_suphor}
\tilde{G}^{(2)}_{\psi}(0,\eta,k) = \frac{i{\Theta(-\eta)}}{2}\Big[e^{ik\eta}(1-ik\eta)-e^{-ik\eta}(1+ik\eta)]\,.
\end{equation}
Taking the super-horizon limit, $k_i\tau \rightarrow 0 $, of \refeq{bk_tau} and substituting \refeq{gpsi_suphor} in it, the bispectrum becomes
\begin{multline}\label{eq:bk_suphor112}
\left\langle \hat{h}_1^R(\tau, \bm k_1)\hat{h}_1^R(\tau, \bm k_2)\hat{h}_2^R(\tau, \bm k_3)\right\rangle  = (2\pi)^3 \delta_D(\bm k_1+\bm k_2+\bm k_3) \Bigg(\frac{H}{M_P}\Bigg)^4 k^2_1\,\tilde{\Xi}^2\,\Xi \frac{2}{(k_1 k_2k_3)^3}
\\
\int_{-\infty}^{0}\, \frac{d\eta}{\eta^2}\,\frac{i}{2}\Big[e^{ik_3\eta}(1-ik_3\eta)-e^{-ik_3\eta}(1+ik_3\eta)]  
{e^{i{(k_1+k_2)}\eta}} [1-{i}{k_1\eta}][1-{i}{k_2\eta}]\,.
\end{multline} 
Following the same procedure as above, we find for the other two terms in \refeq{bk_parts}
\begin{multline}\label{eq:bk_suphor121}
\left\langle \hat{h}_1^R(\tau, \bm k_1)\hat{h}_2^R(\tau, \bm k_2)\hat{h}_1^R(\tau, \bm k_3)\right\rangle  = (2\pi)^3 \delta_D(\bm k_1+\bm k_2+\bm k_3) \Bigg(\frac{H}{M_P}\Bigg)^4 k^2_1\,\tilde{\Xi}^2\,\Xi \frac{2}{(k_1 k_2k_3)^3}
\\
\int_{-\infty}^{0}\,\frac{d\eta}{\eta^2}\,\frac{i}{2}\Big[e^{ik_2\eta}(1-ik_2\eta)-e^{-ik_2\eta}(1+ik_2\eta)]  
{e^{i{(k_1-k_3)}\eta}} [1-{i}{k_1\eta}][1+{i}{k_3\eta}]\,,
\end{multline}
and
\begin{multline}\label{eq:bk_suphor211}
\left\langle \hat{h}_2^R(\tau, \bm k_1)\hat{h}_1^R(\tau, \bm k_2)\hat{h}_1^R(\tau, \bm k_3)\right\rangle  = (2\pi)^3 \delta_D(\bm k_1+\bm k_2+\bm k_3) \Bigg(\frac{H}{M_P}\Bigg)^4 k^2_1\,\tilde{\Xi}^2\,\Xi\frac{2}{(k_1 k_2k_3)^3}
\\
\int_{-\infty}^{0}\, \frac{d\eta}{\eta^2}\,\frac{i}{2}\Big[e^{ik_1\eta}(1-ik_1\eta)-e^{-ik_1\eta}(1+ik_1\eta)]  
{e^{-i{(k_2+k_3)}\eta}} [1+{i}{k_2\eta}][1+{i}{k_3\eta}]\,.
\end{multline}
Adding \refeqs{bk_suphor112}{bk_suphor211}, the first term of \refeq{bk_suphor121} is cancelled by the second term of \refeq{bk_suphor112} while the second term is cancelled by the first term in \refeq{bk_suphor211} so that we find
\begin{multline}\label{eq:bk_final-1}
\left\langle \hat{h}^R(\tau, \bm k_1)\hat{h}^R(\tau, \bm k_2)\hat{h}^R(\tau, \bm k_3)\right\rangle = (2\pi)^3 \delta_D(\bm k_1+\bm k_2+\bm k_3) \Bigg(\frac{H}{M_P}\Bigg)^4 k^2_1\,\tilde{\Xi}^2\,\Xi
\\
\frac{2}{(k_1 k_2k_3)^3}\int_{-\infty}^{0}\, \frac{d\eta}{\eta^2}\,\Big[\frac{i}{2}e^{i(k_1+k_2+k_3)\eta}(1-ik_1\eta)(1-{i}{k_2\eta})(1-{i}{k_3\eta})+c.c.\Big] \,,
\end{multline}
which we can simplify to get
\begin{multline}\label{eq:bk_final}
B^{RRR}_h(k_1, k_2, k_3)=(2\pi)^3 \delta_D(\bm k_1+\bm k_2+\bm k_3) \Bigg(\frac{H}{M_P}\Bigg)^4 k^2_1\,\tilde{\Xi}^2\,\Xi
\\
\frac{2}{(k_1 k_2k_3)^3}\text{Re}\Bigg[\int_{-\infty}^{0}\, i\,\frac{d\eta}{\eta^2}\,e^{i(k_1+k_2+k_3)\eta}(1-ik_1\eta)(1-{i}{k_2\eta})(1-{i}{k_3\eta})\Bigg] \,,
\end{multline}
where $\text{Re}[z]$ denotes the real part of the complex number $z$. This is the same result as obtained via the in-in formalism~\cite{maldacena:2002,maldacena/pimentel:2011,Gao:2011vs}. The integral can be evaluated by choosing the integration contour to go from $-\infty(1+i\epsilon)$ to $0$, with $\epsilon \ll 1$, to get a convergent result~\cite{maldacena:2002,maldacena/pimentel:2011,Gao:2011vs},
\begin{equation}\label{eq:bk_integrated}
(k_1k_2k_3)^2B^{RRR}_h(k_1, k_2, k_3) = (2\pi)^3 \delta_D(\bm k_1+\bm k_2+\bm k_3) 
\Bigg(\frac{H}{M_P}\Bigg)^4 
\frac{2 \,\tilde{\Xi}^2\,\Xi}{ r_2r_3 }\Bigg[\tilde{\Xi}-\frac{\sum_{i>j}r_ir_j}{\tilde{\Xi}}-\frac{r_2r_3}{\tilde{\Xi}^2}\Bigg] \,,
\end{equation}
where $r_i \equiv k_i/k_1$. Note that this choice of the integration contour corresponds to projecting the vacuum of the interacting theory on to that of the free theory in the in-in formalism~\cite{maldacena:2002,Weinberg:2005vy}. The same choice needs to be made also in the Green's function approach. In retrospect, this is not surprising. 

In our approach, the time evolution is carried entirely by the second-order operator, \refeq{psi_2nd_gr}, while states remain stationary. So, the expectation value in \refeq{bk_parts} needs to be evaluated in the vacuum of the interacting theory, the so-called ``in" state. On the other hand, the creation and annihilation operators, $\hat{a}^p_{\bm{k}}$ and $\hat{a}^{\dagger p}_{\bm{k}}$, are  defined in the vacuum of the free theory, where the higher order source terms ($S_{pq}$ with powers of $\psi \geq 2$) in the equation of motion for $\psi_{pq}$ are $0$ over all of spacetime (c.f. \refeq{ann_cre}). As a result, when evaluating the expectation values using $\hat{a}^p_{\bm{k}}$ and $\hat{a}^{\dagger p}_{\bm{k}}$ in \refeq{exp_val}, we need to project the ``in" state on to the vacuum of the free theory. How do we achieve that? Looking at \refeq{psi_2nd_gr} and considering the Heaviside function $\Theta(\tau-\eta)$ in Green's function (\refeq{gpsi}), we see that as $\tau \rightarrow -\infty$ the second-order operator $\rightarrow 0$ (which is equivalent to the interaction Hamiltonian $\rightarrow 0$ in the in-in formalism). Thus, at past infinity the vacuum of the interacting theory is essentially that of the free theory, and the projection prescription then follows from that of the in-in formalism~\cite{maldacena:2002,Weinberg:2005vy}. 

\begin{figure*}
	\centering
	\includegraphics[width=1\textwidth, bb = 0 0 647 406]{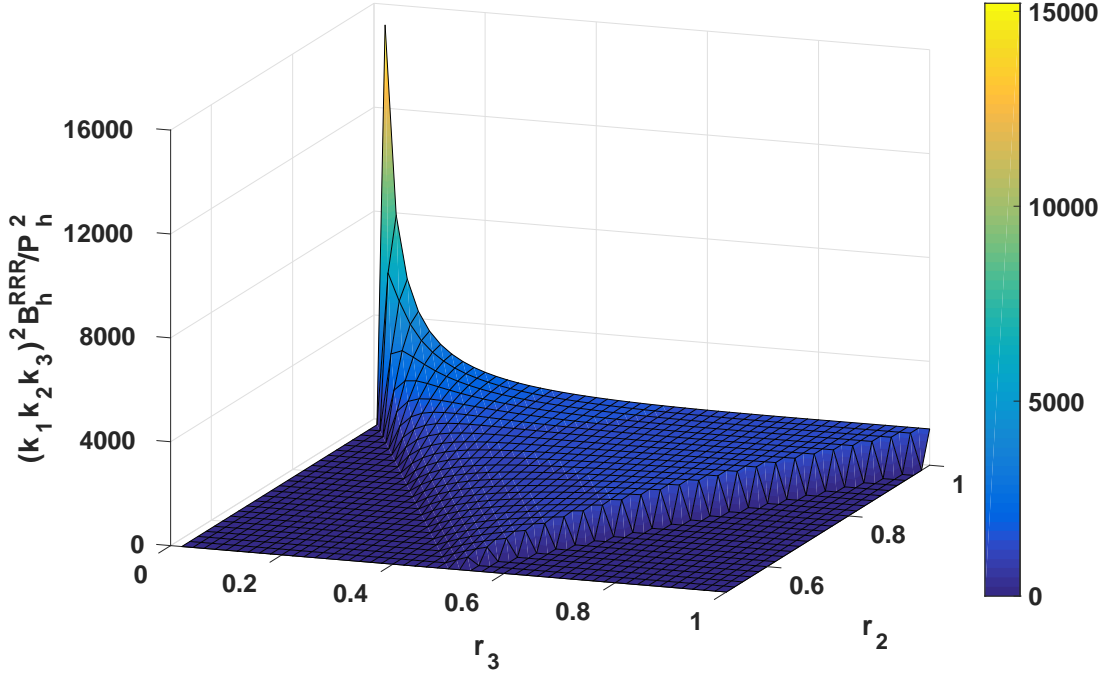}
	\caption{3D plot of ${(k_1k_2k_3)^2B^{RRR}_h}/({\mathcal{P}^R_h})^2$ for right-handed vacuum fluctuations of the metric. We only show $r_3\leq r_2$ and the triangle condition implies that the bispectrum is non-zero only for $r_2+r_3\geq1$. 
	}
	\label{fig:bk_gr}
\end{figure*}

In refs.~\cite{Agrawal:2017awz,Agrawal:2018mrg} right-handed tensor perturbations are chirally amplified by tachyonic instability of an SU(2) gauge field, which leads to highly non-Gaussian GWs. There, non-Gaussianity is characterised using the ratio $B^{RRR}_h(k,k,k)/(P^R_h(k))^2$. We can now evaluate the same ratio for the case of vacuum fluctuations,
\begin{equation}
\frac{B^{RRR}_{h,\text{vac}}(k,k,k)}{(P^{R}_{h,\text{vac}}(k))^2} = 3.586\,,
\end{equation}
which is of order unity and much smaller than that generated in models with spectator gauge  fields~\cite{Agrawal:2017awz,Agrawal:2018mrg,cook/sorbo:2013,barnaby/etal:2012}. \refFig{bk_gr} shows the shape of the tensor bispectrum, \refeq{bk_integrated}, normalised by the square of the dimensionless power spectrum, $(k_1k_2k_3)^2B^{RRR}_h/(\mathcal{P}^R_h)^2$, as a function of $r_2$ and $r_3$. We find that it peaks in the squeezed limit, $r_3 \ll r_2 \approx 1$, and it is zero in the folded limit, $r_2+r_3=1$. 

\subsection{Mixed Bispectrum of Right- and Left-Handed Vacuum Fluctuations}
\label{sec:bispectrumRRL}
Parity invariance means that $B^{RRR}_h = B^{LLL}_h$, $B^{RRL}_h = B^{LLR}_h$, $B^{RLR}_h = B^{LRL}_h$, and $B^{LRR}_h = B^{RLL}_h$, so that one only needs to calculate 4 out of 8 possible bispectra. As a concrete illustration, let us now explicitly evaluate $B^{RRL}_h(k_1, k_2, k_3)$ using Green's functions. We have, analogous to \refeq{bk_parts},
\begin{multline}\label{eq:bkrrl_parts}
\left\langle \hat{\psi}^R(\tau, \bm k_1)\hat{\psi}^R(\tau, \bm k_2)\hat{\psi}^L(\tau, \bm k_3)\right\rangle = \left\langle \hat{\psi}_1^R(\tau, \bm k_1)\hat{\psi}_1^R(\tau, \bm k_2)\hat{\psi}_2^L(\tau, \bm k_3)\right\rangle
\\
+\left\langle \hat{\psi}_1^R(\tau, \bm k_1)\hat{\psi}_2^R(\tau, \bm k_2)\hat{\psi}_1^L(\tau, \bm k_3)\right\rangle+\left\langle \hat{\psi}_2^R(\tau, \bm k_1)\hat{\psi}_1^R(\tau, \bm k_2)\hat{\psi}_1^L(\tau, \bm k_3)\right\rangle\,.
\end{multline}
Now we need to evaluate all three terms separately. The calculation proceeds exactly as before
\begin{multline}\label{eq:bkrrl_raw1}
\left\langle \hat{\psi}_1^R(\tau, \bm k_1)\hat{\psi}_1^R(\tau, \bm k_2)\hat{\psi}_2^L(\tau, \bm k_3)\right\rangle  = \frac{H}{M_P}\int_{-\infty}^{0}\, {d\eta}\,{\eta}\, G_{\psi}^{(2)}(\tau, \eta, k_3)\int\, \frac{d^3p_1 d^3p_2}{(2\pi)^6} \, 
\\
\delta_D(\bm p_1 +\bm p_2-\bm k_3) \,\, Q^{RR}_{pq}(\bm p_1, \bm p_2) e^R_{pq}(\bm k_3) \left\langle \hat{\psi}^R_1(\tau,{\bm{k}_1})\hat{\psi}^R_1(\tau,{\bm{k}_2})\hat{\psi}^R_1(\eta,{\bm{p}_1})\hat{\psi}^R_1(\eta,\bm{p}_2)\right\rangle\,,
\end{multline}
where we only needed $Q^{RR}_{pq}$ because we have two right-handed operators at first-order in the bispectrum. Note that we now use $e^R_{pq}(\textbf{k})$ to calculate the second-order operator for the left-handed polarisation. The expectation value of the first-order fields is given by \refeq{exp_val}. Using that we can write 
\begin{multline}\label{eq:bkrrl_raw1_1}
\left\langle \hat{\psi}_1^R(\tau, \bm k_1)\hat{\psi}_1^R(\tau, \bm k_2)\hat{\psi}_2^L(\tau, \bm k_3)\right\rangle  = \frac{H}{M_P}(2\pi)^3\delta_D(\bm{k}_1+\bm{k}_2+\bm{k}_3)\,
\\ \Big[e^R_{pq}(\bm{k}_3)Q^{RR}_{pq}(-\bm{k}_1,-\bm{k}_2)+e^R_{pq}(\bm{k}_3)Q^{RR}_{pq}(-\bm{k}_2,-\bm{k}_1)\Big]\Psi^R_1(\tau,\bm{k}_1)\Psi^R_1(\tau,\bm{k}_2)
\\
\int_{-\infty}^{0}\, {d\eta}\,{\eta}\, G_{\psi}^{(2)}(\tau, \eta, k_3) \Psi^{*R}_1(\eta,-\bm{k}_1)\Psi^{*R}_1(\eta,-\bm{k}_2)\,.
\end{multline}
For the second term in \refeq{bkrrl_parts} we have
\begin{multline}\label{eq:bkrrl_raw2}
\left\langle \hat{\psi}_1^R(\tau, \bm k_1)\hat{\psi}_2^R(\tau, \bm k_2)\hat{\psi}_1^L(\tau, \bm k_3)\right\rangle  = \frac{H}{M_P}\int_{-\infty}^{0}\, {d\eta}\,{\eta}\, G_{\psi}^{(2)}(\tau, \eta, k_2)\int\, \frac{d^3p_1 d^3p_2}{(2\pi)^6} \, 
\\
\delta_D(\bm p_1 +\bm p_2-\bm k_2) \,\, e^L_{pq}(\bm{k}_2)\Big\langle \hat{\psi}^R_1(\tau,{\bm{k}_1})\Big[Q^{LR}_{pq}(\bm{p}_1,\bm{p}_2)\hat{\psi}^L_1(\eta,{\bm{p}_1})\hat{\psi}^R_1(\eta,\bm{p}_2)
\\
+Q^{RL}_{pq}(\bm{p}_1,\bm{p}_2)\hat{\psi}^R_1(\eta,{\bm{p}_1})\hat{\psi}^L_1(\eta,\bm{p}_2)\Big]\hat{\psi}^L_1(\tau,{\bm{k}_3})\Big\rangle\,,
\end{multline}
where we now use terms containing $Q^{LR}_{pq}$ and $Q^{RL}_{pq}$ in the source function (\refapp{source}) as we have one right-handed and one left-handed mode in our bispectrum. This can be simplified using $\left< \hat{\psi}_1^L \hat{\psi}_1^R \right> = \left< \hat{\psi}_1^R \hat{\psi}_1^L \right> = 0$ to get
\begin{multline}\label{eq:bkrrl_raw2_1}
\left\langle \hat{\psi}_1^R(\tau, \bm k_1)\hat{\psi}_2^R(\tau, \bm k_2)\hat{\psi}_1^L(\tau, \bm k_3)\right\rangle  = \frac{H}{M_P} (2\pi)^3\delta_D(\bm{k}_1+\bm{k}_2+\bm{k}_3)
\\
e^L_{pq}(\bm{k}_2)\Big[Q^{LR}_{pq}(-\bm{k}_3,-\bm{k}_1)+Q^{RL}_{pq}(-\bm{k}_1,-\bm{k}_3)\Big]\Psi^R_1(\tau, \bm{k}_1)\Psi^{*L}_1(\tau, -\bm{k}_3)
\\
\int_{-\infty}^{0}\, {d\eta}\,{\eta}\, G_{\psi}^{(2)}(\tau, \eta, k_2)  \, \Psi^{*R}_1(\eta, -\bm{k}_1)\Psi^L_1(\eta, -\bm{k}_3) \,.
\end{multline}
Similarly, for the third term in \refeq{bkrrl_parts} we get
\begin{multline}\label{eq:bkrrl_raw3}
\left\langle \hat{\psi}_2^R(\tau, \bm k_1)\hat{\psi}_1^R(\tau, \bm k_2)\hat{\psi}_1^L(\tau, \bm k_3)\right\rangle  = \frac{H}{M_P}\int_{-\infty}^{0}\, {d\eta}\,{\eta}\, G_{\psi}^{(2)}(\tau, \eta, k_2)\int\, \frac{d^3p_1 d^3p_2}{(2\pi)^6} \, 
\\
\delta_D(\bm p_1 +\bm p_2-\bm k_2) \,\, e^L_{pq}(\bm{k}_1)\Big\langle \Big[Q^{LR}_{pq}(\bm{p}_1,\bm{p}_2)\hat{\psi}^L_1(\eta,{\bm{p}_1})\hat{\psi}^R_1(\eta,\bm{p}_2)
\\
+Q^{RL}_{pq}(\bm{p}_1,\bm{p}_2)\hat{\psi}^R_1(\eta,{\bm{p}_1})\hat{\psi}^L_1(\eta,\bm{p}_2)\Big]\hat{\psi}^R_1(\tau,{\bm{k}_2})\hat{\psi}^L_1(\tau,{\bm{k}_3})\Big\rangle\,,
\end{multline}
where again we need $Q_{pq}^{LR}$ and $Q^{RL}_{pq}$. This can be simplified to get
\begin{multline}\label{eq:bkrrl_raw3_1}
\left\langle \hat{\psi}_2^R(\tau, \bm k_1)\hat{\psi}_1^R(\tau, \bm k_2)\hat{\psi}_1^L(\tau, \bm k_3)\right\rangle  = \frac{H}{M_P} (2\pi)^3\delta_D(\bm{k}_1+\bm{k}_2+\bm{k}_3)
\\
e^L_{pq}(\bm{k}_1)\Big[Q^{LR}_{pq}(-\bm{k}_3,-\bm{k}_2)+Q^{RL}_{pq}(-\bm{k}_2,-\bm{k}_3)\Big]\Psi^{*R}_1(\tau, -\bm{k}_2)\Psi^{*L}_1(\tau, -\bm{k}_3)
\\
\int_{-\infty}^{0}\, {d\eta}\,{\eta}\, G_{\psi}^{(2)}(\tau, \eta, k_2)  \, \Psi^{R}_1(\eta, \bm{k}_2)\Psi^{L}_1(\eta, \bm{k}_3) \,.
\end{multline}

\begin{figure*}
	\centering
	\includegraphics[width=1\textwidth, bb = 0 0 640 405]{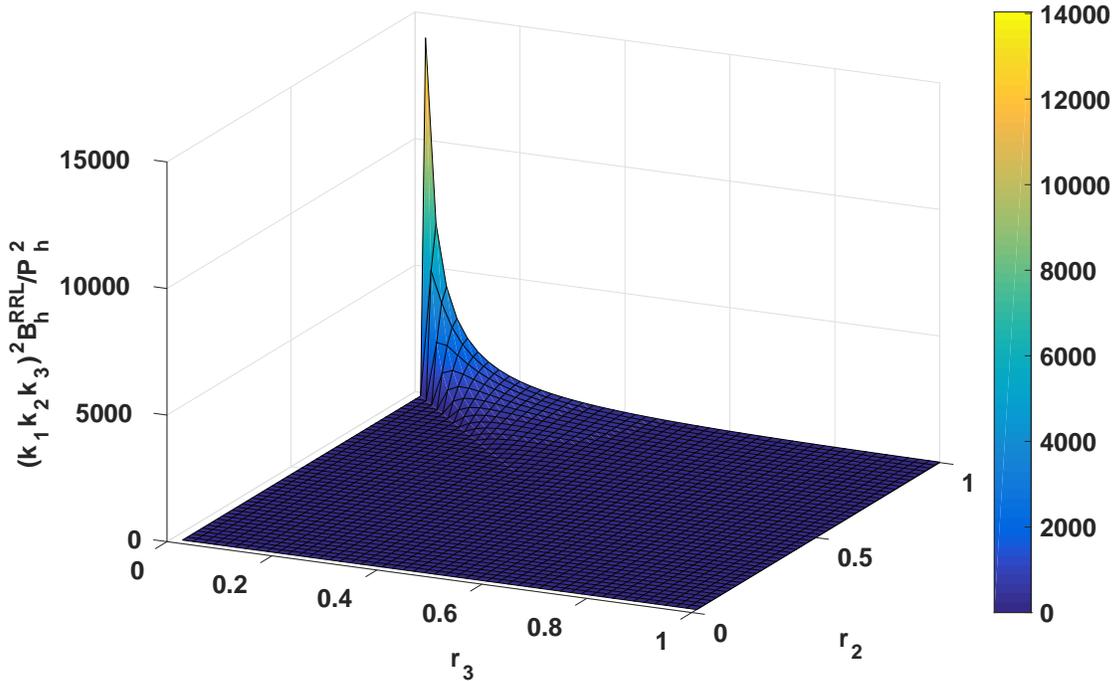}
	\caption{3D plot of ${(k_1k_2k_3)^2B^{RRL}_h}/({\mathcal{P}^R_h})^2$. We do not restrict to $r_3 \leq r_2$ as the bispectrum is not symmetric in $r_2$ and $r_3$. The tensor bispectrum has a sharp peak in the squeezed limit, $r_3 \ll r_2 \approx 1$, but is quite small for most of the $r_3-r_2$ plane.  
	}
	\label{fig:bkRRL_gr}
\end{figure*}

The contraction of polarisation tensors is again found to be the same for all three terms, and is given by
\begin{equation}
k^2_1 \frac{\Xi}{\tilde{\Xi}^2}(1+r_2-r_3)^4\,,
\end{equation}
which can be taken out of all the three terms. The minus sign in front of $r_3$ reflects the breaking of the symmetry inherent in \refeq{bkrrl_parts}. Then, note that the remaining terms containing the first order mode functions are \textit{exactly} the same as in \refsec{bispectrum}, since the first-order mode functions are the same for left- and right-handed polarisations. Thus, we obtain 
\begin{multline}\label{eq:bkrrl_final}
B^{RRL}_h(k_1, k_2, k_3)=(2\pi)^3 \delta_D(\bm k_1+\bm k_2+\bm k_3) \Bigg(\frac{H}{M_P}\Bigg)^4 k^2_1\,\frac{\Xi}{\tilde{\Xi}^2}(1+r_2-r_3)^4
\\
\frac{2}{(k_1 k_2k_3)^3}\text{Re}\Bigg[\int_{-\infty}^{0}\, i\,\frac{d\eta}{\eta^2}\,e^{i(k_1+k_2+k_3)\eta}(1-ik_1\eta)(1-{i}{k_2\eta})(1-{i}{k_3\eta})\Bigg] \,,
\end{multline}
which, after evaluating the integral, yields
\begin{multline}\label{eq:bkrrl_integrated}
(k_1k_2k_3)^2B^{RRL}_h(k_1, k_2, k_3) = (2\pi)^3 \delta_D(\bm k_1+\bm k_2+\bm k_3) 
\Bigg(\frac{H}{M_P}\Bigg)^4 
\\
\frac{2\,\Xi}{\tilde{\Xi}^2}\frac{(1+r_2-r_3)^4}{r_2 r_3}\Bigg[\tilde{\Xi}-\frac{\sum_{i>j}r_ir_j}{\tilde{\Xi}}-\frac{r_2r_3}{\tilde{\Xi}^2}\Bigg] \,,
\end{multline}
which is again the same as that in the in-in formalism~\cite{maldacena:2002,maldacena/pimentel:2011,Gao:2011vs}. \refFig{bkRRL_gr} shows the shape of the mixed bispectrum, \refeq{bkrrl_integrated}, normalised by the power spectrum, as in \reffig{bk_gr}. Note that we show the full range of $0 \leq r_2 \leq 1$ because the bispectrum is no longer symmetric in $r_2$ and $r_3$. While the bispectrum of three right-handed modes has large values over most of the $r_3-r_2$ plane, the mixed bispectrum is much closer to zero for most of the plane, and has a rather sharp peak in the squeezed limit. The ratio of the mixed bispectrum to the ``pure" one can also be calculated (\refeq{bkrrl_integrated} and \refeq{bk_integrated})
\begin{equation}
\frac{B^{RRL}_h(k_1, k_2, k_3)}{B^{RRR}_h(k_1, k_2, k_3)} = \frac{(1+r_2-r_3)^4}{\tilde{\Xi}^4}=\frac{(1+r_2-r_3)^4}{(1+r_2+r_3)^4}\,,
\end{equation}
which in the equilateral configuration, $r_2 = r_3 = 1$, evaluates to $1/81$, and approaches $1$ in the squeezed limit, $r_3  \ll r_2 \approx 1$~\cite{Gao:2011vs}. Note however, that in the opposite squeezed limit, $r_2 \ll r_3 \approx 1$, the bispectrum approaches $0$. This is because when the left-handed mode has a wavelength much larger than the other two, the bispectrum is essentially produced by small-scale self-interactions between two right-handed modes, which are highly correlated. In the other limit, the bispectrum arises from interactions between a left- and right-handed mode, which are much less correlated. We also note that this ratio is close to $0$ for the whole plane in models with spectator gauge fields~\cite{barnaby/etal:2012,cook/sorbo:2013,namba/etal:2015,Agrawal:2017awz,Agrawal:2018mrg}, where tensor fluctuations are dominated by sourced modes of a single helicity, in contrast to vacuum fluctuations. 

\section{Conclusion}
\label{sec:conclusion}

We have calculated the tree-level bispectrum of right-handed vacuum fluctuations of the metric using the Green's function method. The tensor bispectrum from this method agrees with that derived previously using the in-in formalism exactly. We also argued that the same prescription for choosing the integration contour as in the in-in formalism should be used, which amounts to projecting the initial vacuum of the interacting theory on to the free vacuum. Green's function methods are useful for calculating the tensor bispectrum in models where a Lagrangian formulation does not exist, or is unknown. Our results confirm that in this case the answer will be the same as what would have been obtained using the in-in formalism.

\acknowledgments
We would like to thank E. Komatsu for useful discussions and comments on the manuscript, T. Fujita for help with evaluating polarisation tensor contractions, and S. Vadia and T. Lazeyras for comments on the manuscript. 

\appendix
\section{Source Function at Second-Order}
\label{app:source}
In this appendix, we present the source function at second-order, in terms of the contributions from different polarisations of the first-order modes. The source function is given by the right hand side of \refeq{h_eom3},
\begin{multline}\label{eq:source_x}
S_{pq}(\tau, \bm x)=\frac{H}{M_P}\,\tau\,\Big[2\psi_{pl,qr}(\tau, \bm x)\psi_{lr}(\tau, \bm x)+2\psi_{pr,l}(\tau, \bm x)\psi_{ql,r}(\tau, \bm x)
\\
+\psi_{lr,q}(\tau, \bm x)\psi_{lr,p}(\tau, \bm x)-2\psi_{pq,lr}(\tau, \bm x)\psi_{lr}(\tau, \bm x)-2\psi_{lr,q}(\tau, \bm x)\psi_{lp,r}(\tau, \bm x)\Big]\,,
\end{multline}
which in Fourier space gives,
\begin{multline}\label{eq:source_k}
S_{pq}(\tau, \bm k)=\frac{H}{M_P}\,\tau\,\int \frac{d^3 p_1 \,d^3p_2}{(2\pi)^6}\delta_D(\bm{p}_1+\bm{p}_2-\bm{k})\Big[2p_{1q}p_{1r}\psi_{pl}(\tau, \bm p_1)\psi_{lr}(\tau, \bm p_2)
\\
+2p_{1l}p_{2r}\psi_{pr}(\tau, \bm p_1)\psi_{ql}(\tau, \bm p_2)+p_{1q}p_{2p}\psi_{lr}(\tau, \bm p_1)\psi_{lr}(\tau, \bm p_2)
\\
-2p_{1l}p_{1r}\psi_{pq}(\tau, \bm p_1)\psi_{lr}(\tau, \bm p_2)-2p_{1q}p_{2r}\psi_{lr}(\tau, \bm p_1)\psi_{lp}(\tau, \bm p_2)\Big]\,,
\end{multline}
where we have defined (c.f. \refeq{psi_decomp})
\begin{equation}
\psi_{ij}(\tau, \bm{p}_n) \equiv \Big[e^R_{ij}(\bm{p}_n)\psi_{\bm{p}_n}^R(\tau)+e^L_{ij}(\bm{p}_n)\psi_{\bm{p}_n}^L(\tau)\Big] = \sum_{c = L,R} e^c_{ij}(\bm{p}_n)\psi_{\bm{p}_n}^c(\tau)\,,
\end{equation}
with $n=1$ or $2$. Substituting this in \refeq{source_k} we find
\begin{equation}\label{eq:source_k_parts}
S_{pq}(\tau, \bm k)=\frac{H}{M_P}\,\tau\, \int \frac{d^3 p_1 \,d^3p_2}{(2\pi)^6}\delta_D(\bm{p}_1+\bm{p}_2-\bm{k}) \Big[S^{LL}_{pq}+S^{LR}_{pq}+S^{RL}_{pq}+S^{RR}_{pq}\Big]\,,
\end{equation}
with the different parts given as
\begin{align}\label{eq:parts_of_source}
\nonumber S^{LL}_{pq} &= \psi^L_{\bm{p}_1}\psi^L_{\bm{p}_2}\Big[2e^L_{pl}(p_1)e^L_{lr}(p_2)p_{1q}p_{1r}+2e^L_{pr}(p_1)e^L_{ql}(p_2)p_{1l}p_{2r}+e^L_{rl}(p_1)e^L_{rl}(p_2)p_{1p}p_{2q}
\\
&-2e^L_{pq}(p_1)e^L_{lr}(p_2)p_{1l}p_{1r}-2e^L_{lr}(p_1)e^L_{lp}(p_2)p_{1q}p_{2r}\Big]\equiv \psi^L_{\bm{p}_1}\psi^L_{\bm{p}_2}Q^{LL}_{pq}\,,
\\
\nonumber S^{LR}_{pq} &= \psi^L_{\bm{p}_1}\psi^R_{\bm{p}_2}\Big[2e^L_{pl}(p_1)e^R_{lr}(p_2)p_{1q}p_{1r}+2e^L_{pr}(p_1)e^R_{ql}(p_2)p_{1l}p_{2r}+e^L_{rl}(p_1)e^R_{rl}(p_2)p_{1p}p_{2q}
\\
&-2e^L_{pq}(p_1)e^R_{lr}(p_2)p_{1l}p_{1r}-2e^L_{lr}(p_1)e^R_{lp}(p_2)p_{1q}p_{2r}\Big]\equiv\psi^L_{\bm{p}_1}\psi^R_{\bm{p}_2}Q^{LR}_{pq}\,,
\\
\nonumber S^{RL}_{pq} &= \psi^R_{\bm{p}_1}\psi^L_{\bm{p}_2}\Big[2e^R_{pl}(p_1)e^L_{lr}(p_2)p_{1q}p_{1r}+2e^R_{pr}(p_1)e^L_{ql}(p_2)p_{1l}p_{2r}+e^R_{rl}(p_1)e^L_{rl}(p_2)p_{1p}p_{2q}
\\
&-2e^R_{pq}(p_1)e^L_{lr}(p_2)p_{1l}p_{1r}-2e^R_{lr}(p_1)e^L_{lp}(p_2)p_{1q}p_{2r}\Big]\equiv\psi^R_{\bm{p}_1}\psi^L_{\bm{p}_2}Q^{RL}_{pq}\,,
\\
\nonumber S^{RR}_{pq} &= \psi^R_{\bm{p}_1}\psi^R_{\bm{p}_2}\Big[2e^R_{pl}(p_1)e^R_{lr}(p_2)p_{1q}p_{1r}+2e^R_{pr}(p_1)e^R_{ql}(p_2)p_{1l}p_{2r}+e^R_{rl}(p_1)e^R_{rl}(p_2)p_{1p}p_{2q}
\\
&-2e^R_{pq}(p_1)e^R_{lr}(p_2)p_{1l}p_{1r}-2e^R_{lr}(p_1)e^R_{lp}(p_2)p_{1q}p_{2r}\Big]\equiv\psi^R_{\bm{p}_1}\psi^R_{\bm{p}_2}Q^{RR}_{pq} \,,
\end{align}
where we have separated the factors containing the polarisation tensors as $Q^{c_1c_2}_{pq}$, since they are not quantized. Then, in order to obtain correlation functions of GWs generated from quantum fluctuations, we quantize the fields $\psi_{\bm{p}}^c(\tau) \rightarrow \hat{\psi}_1^c(\tau, \bm{p})$, and use the condition of Gaussianity on $\hat{\psi}_1$ so that Wick's theorem can be used, and that at first-order, different polarisations are uncorrelated, $\left< \hat{\psi}_1^L \hat{\psi}_1^R \right> = \left< \hat{\psi}_1^R \hat{\psi}_1^L \right> = 0$. As a consequence of this, we see that whenever we want to evaluate a bispectrum with two other first-order modes, the only non-zero contributions arise from parts of $S_{pq}$ containing the same number of left- and right-handed modes as the other two.  

\bibliography{references}
\end{document}